\documentclass{appolb}
\usepackage{epsfig}

\begin{document}
\pagestyle{plain}
\newcount\eLiNe\eLiNe=\inputlineno\advance\eLiNe by -1
\title{SLEPTON FLAVOUR VIOLATION AT COLLIDERS}
\author{Jan KALINOWSKI
\address{Instytut Fizyki Teoretycznej, Uniwersytet Warszawski,
Ho\.za~69,  00-681~Warszawa,  Poland\\
\vspace{1cm}
{\it Dedicated to Stefan Pokorski on the occasion of his 60th birthday}
}}
\maketitle

\begin{abstract}
In supersymmetric extensions of the Standard Model, 
the lepton flavour violation (LFV) is closely related to the
structure of slepton masses and mixing. 
Allowing for the most general flavour structure of the slepton sector, 
consistent with the experimental limits on rare lepton decays,   
large and distinct signals of LFV at future
colliders   can be expected. A case study of mixing of second and third
generation of sleptons at an $e^+e^-$ collider is presented and
compared to that of $\tau\to \mu\gamma$ rare decay.


\end{abstract}

Observations of flavour changing neutral current processes provide
important tests of physics beyond the Standard Model. It is well known
that in the Standard Model the renormalizability, Lorentz and gauge
invariance  
force the  individual lepton
flavour numbers $L_e$, ~$L_\mu$ and $L_\tau$ to be conserved in
addition to the conserved baryon $B$ and total lepton $L$
numbers. These conservation laws are consequences of global symmetries
which are ``accidental'' in the sense that they
follow from the spin and gauge quantum number assignments of the SM
fields.

Experiments on solar and atmospheric neutrinos \cite{neutrino}
provide a compelling evidence for oscillations among three active
neutrinos with different masses. This phenomenon is lepton flavour
violating  and it
is a first direct evidence for physics beyond the Standard Model.
The most favoured model to account for the neutrino masses and
their oscillations is the seesaw mechanism \cite{seesaw} 
with heavy right-handed neutrinos $N$.
The smallness of $m_{\nu_i}$ is obtained in a natural way if the
masses of right-handed neutrinos are assumed in the range $M_N \sim
10^{13}-10^{15}$ GeV, and non-diagonal elements of the Yukawa
couplings of $N$ and $\nu$ generate neutrino mixing.  Radiative
corrections due to these couplings also induce flavour mixing in the
charged lepton sector. An interesting question then arises whether
processes with charged-lepton flavour violation, like $\mu\rightarrow
e\gamma$, $\tau\rightarrow\mu\gamma$ etc., can be generated at
observable rates \cite{LMS}.

In the Standard Model with right-handed neutrinos the charged LFV
decays are strongly suppressed \cite{muegsm} 
via the GIM mechanism ($\sim \Delta
m_\nu^4/M_W^4$).  
In the supersymmetric
extension of this model, however,
the situation of LFV processes may be quite different.
In addition to the seesaw mechanism, new
sources of flavour violation in the leptonic sector can be generated
by soft supersymmetry breaking terms, e.g. 
\begin{eqnarray}
{\cal L}_{\rm soft} \ni  m^2_{L\alpha\beta} \tilde{e}^*_\alpha
\tilde{e}_\beta 
+m^2_{R\alpha\beta}\tilde{e}^*_\alpha\tilde{e}_\beta 
+( A_{\alpha\beta}\tilde{e}^*_\alpha h^0_1\tilde{e}_\beta +
{\mbox{h.c.}}) \label{soft} 
\end{eqnarray}
where only scalar mass and trilinear terms in the leptonic sector have
been written explicitly using self-explanatory notation, $\alpha,
\beta=e,\mu, \tau$.  The
trilinear term, after electroweak symmetry breaking, couples left- and
right-handed charged sleptons through the mass matrix ${m}^2_{LR\alpha\beta}$
which receives a contribution from $ A_{\alpha\beta} \langle h^0_1 \rangle$.  
In general the slepton mass matrix need
not simultaneously be diagonalized with leptons.  If we now rotate
sleptons to the mass eigenstate basis, $\tilde{e}_i=W_{i\alpha}\, \tilde{e}_\alpha$, the slepton-mass
diagonalization matrix $W_{i\alpha}$ enters the chargino and neutralino
couplings
\begin{eqnarray}
\tilde{e}_{i} {W^*}_{i\alpha} \bar{e}_{\alpha} \tilde{\chi}^0 
+\tilde{\nu}_{i} {W^*}_{i\alpha} \bar{e}_{\alpha} \tilde{\chi}^- 
+\ldots 
\end{eqnarray}
and mixes lepton flavour (Latin and Greek subscripts are
slepton mass-eigenstate and flavour 
indices, respectively). 
Contributions form virtual slepton exchanges can therefore 
enhance the rates of rare decays, like $\mu\to e\gamma$. 
Although these contributions are
suppressed through the superGIM mechanism by $\Delta
m_{\tilde{l}}/\bar{m}_{\tilde{l}}$ with the mass difference $\Delta
m_{\tilde{l}}$ and the average mass $\bar{m}_{\tilde{l}}$ of the
sleptons, 
the present experimental upper limits on these processes
\cite{exp} impose already strong bounds on LFV sources in the slepton sector,
in particular for the first two generations of sleptons.

Even if the slepton mass matrix is assumed to be flavour conserving at tree
level to avoid the supersymmetric flavour-changing problem, like
in minimal supergravity or gauge mediated SUSY breaking models, the
off-diagonal terms can be induced radiatively in the framework of the
seesaw mechanism. The reason is that non-diagonal neutrino
mass terms originating from the lepton Yukawa coupling contribute to
the renormalization-group running of $m^2_{Lij}$, $m^2_{Rij}$ and
$A_{ij}$ matrices \cite{rad}, inducing flavour-mixing entries. 

In extended models, however, additional 
off-diagonal entries are in general  generated. 
For example, in  models with quarks and leptons unified in 
larger multiplets the non-diagonal terms are generated  radiatively by the 
top quark Yukawa
couplings \cite{BH}.  Also string-inspired models naturally lead to
non-universal soft-SUSY breaking terms \cite{string}.  
Flavour changing slepton
exchanges originating from these additional terms can significantly
contribute to
neutrino masses and mixings linking, for example,  
substantial $\nu_\mu - \nu_\tau$ mixing  
with   large $\tilde\mu_L - \tilde\tau_L$ and
$\tilde\nu_\mu - \tilde\nu_\tau$ mixings.  
It is an interesting and open
question whether these terms are required to account for the observed
pattern of  neutrino masses and mixings \cite{CS}.

Once superpartners are discovered, it will be possible to probe lepton
flavour violation directly in their production and decay at future
colliders.  A flavour-violating signal is obtained from the production
of real sleptons (either directly or from chain decays of other
sparticles), followed by their subsequent decays into different flavour
leptons, with missing energy and jets in the final state.   
Searches for these signals have a number of
advantages. First, once kinematically accessible, superpartners are
produced with large cross sections.  Second, flavour changing decays
of sleptons occur at tree level while rare radiative decays of leptons
at one-loop.  Third, they are suppressed only as $\Delta
m_{\tilde{l}}/\Gamma_{\tilde{l}}$ \cite{feng} in contrast to the
$\Delta m_{\tilde{l}}/m_{\tilde{l}}$ suppression of radiative decays -
an important difference since $m_{\tilde{l}}/\Gamma_{\tilde{l}}$ is
typically of the order $10^2$--$10^3$. As a result, allowing for the
most general slepton mass matrix respecting present bounds on rare
lepton decays, large LFV signals are expected both at the LHC
\cite{LHC} and $e^+e^-$ colliders
\cite{feng,ACFH205,nojiri,GKR,PM}. All the above features   
suggest that future $e^+e^-$ colliders (and also the LHC in some
favourable cases) may provide a more powerful tool to search for and
explore supersymmetric lepton flavour violation than rare decay
processes.

In this work we concentrate on the question how well models of LFV can
be probed at future $e^+e^-$ colliders. As a case study we consider a
pure 2-3 intergeneration mixing between $\tilde\nu_\mu$ and
$\tilde\nu_\tau$, generated by a near-maximal mixing angle
$\theta_{23}$, and ignore any mixings with $\tilde\nu_e$. Our work is
closely related to, and extension of, Ref.\cite{GKR} by comparing the
expected reach at the $e^+e^-$ collider to that from the rare decay process
$\tau\to\mu\gamma$.

The
scalar neutrino mass matrix $M^2_{\tilde\nu}$, restricted to the 2-3
generation subspace, can be written in the flavour basis
as
\begin{equation}
{M}^2_{\tilde\nu} = \left(\matrix{\cos\theta_{23} & -\sin\theta_{23} \cr
\sin\theta_{23} & \cos\theta_{23}}\right) 
\left(\matrix{{m}_{\tilde\nu_2} & 0 \cr 0 &
{m}_{\tilde\nu_3}}\right) 
\left(\matrix{\cos\theta_{23} & \sin\theta_{23} \cr
-\sin\theta_{23} & \cos\theta_{23}}\right)
\label{one}
\end{equation}
where ${m}_{\tilde\nu_2}$ and ${m}_{\tilde\nu_3}$ are the
physical masses 
of $\tilde\nu_2$ and $\tilde\nu_3$ respectively.  Its off-diagonal
element is related to physical
masses and mixing angle by 
\begin{equation}
(M^2_{\tilde{\nu}})_{\mu\tau}=\frac{1}{2}(m^2_{\tilde\nu_2}-
m^2_{\tilde\nu_3})\sin2\theta_{23} \label{offdiag}
\end{equation}
In the following we
take the mixing angle $\theta_{23}$ and $\Delta m_{23} =
|{m}_{\tilde\nu_2} - {m}_{\tilde\nu_3}|$ as free, independent
parameters. The same goes for the charged slepton sector, modulo
standard LR mixing, where $\theta_{23}$ and $\Delta m_{23}$ are then
the corresponding parameters for charged sleptons. 

In discussing supersymmetric LFV collider signals one has to consider
two cases in which 
oscillation of lepton flavour can occur in processes with single
(uncorrelated) or 
correlated slepton pair production. The difference comes from the
quantum interference between production and decay \cite{feng}.

Uncorrelated sleptons  may be produced in
cascade decays of heavier nonleptonic superparticles. Such processes are
particularly important for
hadron colliders, where sleptons can be products of uncorrelated
decays of  gluinos or
squarks, but they may also be relevant for lepton colliders where
single slepton can be a decay product of a chargino or neutralino. 
The cross section for the process
\begin{eqnarray}
f\,f'\to e^+_\alpha \, X\,  
\tilde{e}^-_i
\to e^+_\alpha \, X\, e^-_\beta\, Y
\end{eqnarray}   
assuming  negligible generation dependence, nearly degenerate
in mass and narrow sleptons, $\Delta m^2_{ij}$, 
$m\Gamma << m^2$,  
and the case of 
2-3 intergeneration  mixing, takes the simple form \cite{jk}     
\begin{eqnarray}
\sigma_{\alpha\beta}&=&\chi_{23} \sin^2 2\theta_{23}\;
\sigma(f\,f'\to e^+_\alpha \, X\, \tilde{e}^-_\alpha)\nonumber\\[2mm]
&&\times BR( \tilde{e}^-_\alpha \to e^-_\alpha\, Y) \\
\chi_{23} &=& \frac{x_{23}^2}{2(1+x_{23}^2)}, \qquad 
x_{23} = \Delta m_{23}/\Gamma
\end{eqnarray}

Correlated slepton pair production is the dominant slepton production
mechanism at lepton colliders, but it may also occur at hadron
colliders when sleptons are produced in the 
Drell-Yan process. Assuming only the $s$-channel
production mechanism and the same approximations as in
the previous case, the cross section for the process  
\begin{eqnarray}
\bar f\,f\to \tilde{e}^+_i \, \tilde{e}^-_i
\to e^+_\alpha \, X\, e^-_\beta\, Y
\end{eqnarray}
can be written as 
\begin{eqnarray} 
\sigma_{\alpha\beta}&=&\chi_{23}(3-4 \chi_{23})  
\sin^2 2\theta_{23} \;
\sigma(\bar f\,f\to \tilde{e}^+_\alpha \,
\tilde{e}^-_\alpha) \nonumber \\[2mm]
&&\times BR(\tilde{e}^+_\alpha \to e^+_\alpha \, X) \times 
BR(\tilde{e}^-_\alpha \to  e^-_\alpha\, Y)
\end{eqnarray}

In the limit $x_{23}\gg 1$, $\chi_{23} $ approaches 1/2, the
interference can be neglected and the cross section behaves as $\sigma
\sim \sin^22\theta_{23}$ . In the opposite case, the interference
suppresses the flavour changing process, $\sigma \sim (\Delta m_{23}
\sin2\theta_{23}/\Gamma)^2$. The effect of the factor $\chi_{23}$
determines the 
characteristic features of the contour lines of the constant cross
sections in the plane $\Delta m_{23}-\sin2\theta_{23}$, which are  
also visible in Figure 1, where the results of our analysis are
shown.

Our analysis for the LFV signal and background at a 500 GeV $e^+e^-$
linear collider 
has been performed for one of the MSSM representative points  
chosen for detailed case studies at the ECFA/DESY \cite{tesla}. 
This point is given in terms of a mSUGRA
scenario defined by:
 $m_0=100$ GeV, $M_{1/2}=200$ GeV, $A_0=0$ GeV, $\tan\beta=3$ and 
${\rm sgn}(\mu)=+$.  
The corresponding masses of chargino,
neutralino and slepton states, along with some branching ratios,   
relevant for the LFV processes at
$\sqrt{s}=500$ GeV  
are shown in Table 1.

\begin{table}[!t]
\begin{center}
\begin{tabular}{|c|c|c|c||c|c|c|c|}
\hline
 & Mass & Decay  & BR &  & Mass & Decay  & BR\\
\hline
$\tilde\chi^+_1$ & 128 &$\tilde\chi^0_1 q \bar{q}'$ & 0.56 
&$\tilde\chi^0_1$ & 72  &&\\
$\tilde\chi^+_2$ & 346 & $\ell^+\tilde{\nu}_\ell$ & 0.03 $\times$ 3  
&$\tilde\chi^0_2$ & 130 &  $\tilde\chi^0_1 \ell^+\ell^-$ & 0.64  \\
$\tilde{\ell}^-_L$ & 176 & $\tilde{\chi}^-_1\nu_{\ell}$& 0.53
&$\tilde{\nu}_{\ell}$ & 161 &$\tilde{\chi}^+_1{\ell}^-$& 0.48\\
$\tilde{\tau}^-_1$ & 131 & $\tilde{\chi}^0_1 \tau^-$& 1.00&
$\tilde{\tau}^-_2$ & 177 &$\tilde{\chi}^0_i \tau^- $& 0.47\\
\hline
\end{tabular}
\end{center}
\caption{\small The masses (in GeV) and the branching ratios for
decay modes  of light supersymmetric particles which are
relevant to our study. No slepton mixing is assumed.  
$\ell$ denotes $e$ or $\mu$, and $\tau$ unless the 
entry for $\tau$ is explicitly shown.   }
\end{table}

The LFV signal comes from the following processes ($i=2,3$)
\begin{eqnarray}
e^+e^- & \rightarrow &
\tilde{\ell}^-_i\tilde{\ell}^+_i \rightarrow  \tau^+\mu^- \tilde{\chi}^0_1
\tilde{\chi}^0_1, \label{slelfv} \\
e^+e^- & \rightarrow &
\tilde{\nu}_i\tilde{\nu}^c_i  \rightarrow  \tau^+\mu^- \tilde{\chi}^+_1
\tilde{\chi}^-_1 \label{snulfv}\\
e^+e^- & \rightarrow &
\tilde\chi^+_2\tilde{\chi}^-_1   \rightarrow  \tau^+\mu^- 
 \tilde\chi^+_1\tilde\chi^-_1  \label{charlfv}\\
e^+e^- & \rightarrow &
\tilde\chi^0_2  \tilde{\chi}^0_1 \rightarrow  \tau^+\mu^-
\tilde\chi^0_1\tilde\chi^0_1 \label{neulfv}
\end{eqnarray}
where $\tilde{\chi}^\pm_1 \rightarrow \tilde{\chi}^0_1 f\bar{f}'$, and
$\tilde{\chi}^0_1$ escapes detection.  The signature therefore would
be $\tau^{\pm}\mu^{\mp}+
\mbox{4 jets}+ {E\!\!\!/}_T$,  $\tau^{\pm}\mu^{\mp}+ 
\ell + \mbox{2 jets}+ {E\!\!\!/}_T$, or $\tau^{\pm}\mu^{\mp}+ {E\!\!\!/}_T$, 
depending on hadronic or leptonic
$\tilde{\chi}^\pm_1$ decay mode.
If both charginos are required to decay hadronically, the signal 
$\tau^{\pm}\mu^{\mp}+ 4\mbox{ jets}+ {E\!\!\!/}_T$ comes from
(\ref{snulfv}), (\ref{charlfv}) and (\ref{neulfv})
and is SM-background free. The flavour-conserving processes analogous
to (\ref{snulfv}--\ref{neulfv}), but with two $\tau$'s in the final
state where  one of the $\tau$'s
decays leptonically to $\mu$, contribute to the
background. On the other hand, if 
jets are allowed to overlap,    
an important SM background to the final states with $\tau^\pm\mu^\mp + \ge 3
jets+ {E\!\!\!/}_T$ comes from 
$e^+ e^- \rightarrow t \bar t g$. 
\begin{figure}
 \begin{center}
\epsfig{figure=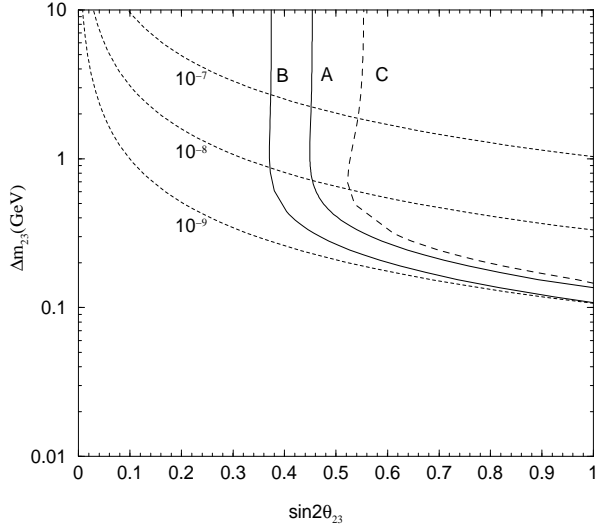,width=8cm,height=7cm}
 \end{center}
\caption[{\bf Figure 1:}]
{\it The 3$\sigma$ significance contours (for the SUSY point mentioned in
 the text) in $\Delta m_{23} -\sin2\theta_{23}$ plane for
 $\sqrt{s}=$500 GeV and for different luminosity options, contours A
 and 
 B being for 500 fb$^{-1}$ and 1000 fb$^{-1}$,
 respectively. The dashed line is for only $\tilde\nu \tilde\nu^c$
 contribution with luminosity 500 fb$^{-1}$.  The dotted 
 lines show contours for BR($\tau\to\mu\gamma$)=10$^{-7}$, 10$^{-8}$
 and 10$^{-9}$.  }
 \label{fig1}
\end{figure}

The results of a simple parton level simulation with a number of  kinematic
cuts listed in \cite{GKR} is shown in Fig.1. 
The significance is given by $\sigma_d =
\frac{N}{\sqrt{N+B}}$ where N and B is the number of signal and
background events respectively for a given luminosity. Fig.1 shows 
the region (to the right of the curve) in the $\Delta
m_{23} - \sin2\theta_{23}$ plane that can be explored or ruled out at
a 3$\sigma$ level by the linear collider of energy 500 GeV for the
given integrated luminosity. The contour (A) is for 500
fb$^{-1}$ and  (B) for 1000
fb$^{-1}$, whereas the dashed line (C) shows the reach of the process 
$\tilde{\nu}_i\tilde{\nu}^c_i$ alone (which were previously studied in 
\cite{feng,nojiri}) 
using our cuts and 
assuming luminosity of  500 fb$^{-1}$. Comparing the dashed line with line 
(A) it has been concluded in Ref.\cite{GKR}  that the chargino
contribution 
increases the sensitivity range to $\sin^2\theta_{23}$ by 10-20\% while 
the sensitivity to $\Delta m_{{23}}$ does not change appreciably.

In the same figure the contour lines for constant branching ratios of
$\tau\to \mu\gamma$ are shown for comparison. 
In the limit of small mass
splitting, the BR($\tau\to\mu\gamma$) can be calculated in the flavour
basis using the mass
insertion technique \cite{gabbiani}.   The LFV charged lepton radiative 
decay takes place through one or more slepton mass insertions, each
bringing a factor of
$\delta^{MN}_{\alpha\beta}=  (M^2_{\tilde{l}})^{MN}_{\alpha\beta}
/\bar{m}^2_{\tilde{l}}$, where $(M^2_{\tilde{l}})^{MN}_{\alpha\beta}$
with $M,N=L,R$ are off-diagonal elements of the slepton mass matrix. 
In our 2-3 intergeneration mixing and using (\ref{offdiag}), the  
radiative process $\tau\to\mu\gamma$  therefore 
constrains    
\begin{equation}
\delta_{\mu\tau}= \frac{\Delta m_{23}}{\bar{m}_\nu}\sin2\theta_{23}
\end{equation}
The contours in Fig.1 have been obtained from the approximate formula
of Ref.\cite{FNS}, normalized to the current experimental limit,
\begin{equation}
BR(\tau\to\mu\gamma)\sim 1.1 \times 10^{-6} \mbox{ 
max}[(\frac{\delta^{LL}_{\mu\tau}}{1.4})^2,
(\frac{\delta^{LR}_{\mu\tau}}{8.3\times 10^{-3}})^2](\frac{100 \mbox{
GeV}}{ \bar{m}_\nu})^4 \label{max}
\end{equation}
which serves only as an order of magnitude estimate of an upper limit
for the supersymmetric contribution to the radiative lepton decay
corresponding to the point in the $\Delta m_{23}-\sin2 \theta_{23}$
plane.  The exact result, which is sensitive to the details of mass
spectra and mixings, can in fact be much smaller due to cancellations
among different contributions.  Nevertheless, even if cancellations do
not occur, Fig.1 demonstrates  that information from the
slepton production and decay is very competitive and, in particular,  
can help to explore
small $\Delta m_{23}$ region.\\[2mm]

To conclude: 
If superpartners are discovered, lepton flavour violating processes
can be observed in slepton production and decay processes at future
colliders. We have demonstrated that their analysis at $e^+e^-$
collisions provides an opportunity to look for $\tau$-$\mu$ flavour
violation and that it is largely complementary to the search for
$\tau\to\mu\gamma$. Although the discussion  has been done in a specific
model, the analysis has  been phrased in as model-independent way as
possible, and can easily be applied to cases with different slepton
mixing patterns. The observation (or non-observation) 
of such processes would provide
important clues about the flavour structure.

\bigskip
\noindent {\bf Acknowledgements:}  The work was supported by the KBN
Grant  5 P03B 119 20 (2001-2002).

\end{document}